\documentclass[jap,twocolumn, showpacs,preprintnumbers,amsmath,amssymb,superscriptaddress,floatfix]{revtex4} 
\usepackage[]{graphicx}
\usepackage{graphics}
\usepackage{subfigure}
\usepackage{amsmath}
\usepackage{tabularx}
\usepackage{bm}
\usepackage{multirow}

\begin{document}

\title[Boosting the Figure Of Merit of LSPR-based refractive index sensing by phase-sensitive measurements]{Boosting the Figure Of Merit of LSPR-based refractive index sensing by phase-sensitive measurements}

\author{Kristof Lodewijks}
\email[Corresponding author: ] {kristof.lodewijks@imec.be}
\affiliation{IMEC, Kapeldreef 75, B-3001 Leuven, Belgium}
\affiliation{Department of electrical engineering (ESAT), KU Leuven, Leuven, Belgium}

\author{Willem Van Roy}
\affiliation{IMEC, Kapeldreef 75, B-3001 Leuven, Belgium}

\author{Gustaaf Borghs}
\affiliation{IMEC, Kapeldreef 75, B-3001 Leuven, Belgium}
\affiliation{Department of Physics and Astronomy, KU Leuven, Leuven, Belgium}

\author{Liesbet Lagae}
\affiliation{IMEC, Kapeldreef 75, B-3001 Leuven, Belgium}
\affiliation{Department of Physics and Astronomy, KU Leuven, Leuven, Belgium}

\author{Pol Van Dorpe}
\affiliation{IMEC, Kapeldreef 75, B-3001 Leuven, Belgium}
\affiliation{Department of electrical engineering (ESAT), KU Leuven, Leuven, Belgium}

\begin{abstract}
Localized surface plasmon resonances possess very interesting properties for a wide variety of sensing applications. In many of the existing applications only the intensity of the reflected or transmitted signals is taken into account, while the phase information is ignored. At the center frequency of a (localized) surface plasmon resonance, the electron cloud makes the transition between in- and out-of-phase oscillation with respect to the incident wave. Here we show that this information can experimentally be extracted by performing phase-sensitive measurements, which result in linewidths that are almost one order of magnitude smaller than those for intensity based measurements. As this phase transition is an intrinsic property of a plasmon resonance, this opens up many possibilities for boosting the figure of merit (FOM) of refractive index sensing by taking into account the phase of the plasmon resonance. We experimentally investigated this for two model systems: randomly distributed gold nanodisks and gold nanorings on top of a continuous gold layer and a dielectric spacer and observed FOM values up to 8.3 and 16.5 for the respective nanoparticles. 
\end{abstract}

\maketitle

The collective oscillations of the free electrons in nobal metals have been studied extensively over the past decades for a wide variety of applications. Both propagating surface plasmon polariton (SPP) and localized surface plasmon resonance (LSPR) modes posess very interesting properties, with applications in sensing (refractive index sensing \cite{ref:1,ref:2,ref:3,ref:4,ref:5,ref:6}, SERS \cite{ref:7}), metamaterials \cite{ref:8}, waveguiding \cite{ref:9, ref:10} and enhanced coupling to active semiconductor components (e.g. photovoltaic cells \cite{ref:11}, SPASERs \cite{ref:12, ref:13}). Refractive index sensing is by far the most studied application and allows label-free and real-time detection of changes in the dielectric environment of the plasmonic nanostructures. Moreover, by functionalizing the nanostructures, the sensors can be made specific to a particular molecule. In former works, research groups have followed various paths in order to optimize the nanostructure designs such that higher sensitivities ($d\lambda/dn$) figures of merit ($FOM = (d\lambda/dn)/fwhm$) and lower detection limits (DLs) can be achieved. Recently a lot of progress has been made in line width tuning (Fano resonances, sub-radiance) to reach this goal \cite{ref:1, ref:2, ref:14, ref:15, ref:16, ref:17}. 
In any resonant system, a pronounced transition from in- to out-of-phase oscillation is observed around the center frequency of the resonance with respect to the driving force. This is also the case for (localized) surface plasmon resonances, where the electron cloud makes the transition between in- and out-of-phase oscillation with respect to the incident wave. In case of conventional SPR sensing, it was already shown that these transitions can be probed by phase sensitive measurements, which show a much smaller spectral/angular footprint compared to their intensity based counterparts \cite{ref:3,ref:4}. Here we show that using standard spectroscopic ellipsometry measurements, we can measure similar phase jumps around the center frequency of localized surface plasmon resonances. For our two model systems, we investigated the angle- and polarization dependent reflection spectra and the phase difference between P- and S-polarized waves, using lock-in measurements. 

In this work we investigated localized surface plasmon resonances in randomly distributed gold nanoparticles on top of a continuous gold layer and a dielectric spacer by spectroscopic ellipsometry. The investigated sample structures are illustrated in Fig. 1 and consist of a glass substrate covered with a $100 nm$ gold (Au) layer, a $50 nm$ silica ($SiO_2$) spacer and randomly distributed Au nanoparticles that were fabricated using colloidal lithography \cite{ref:18, ref:19}. The nanorings (Fig 1 a and c) have an outer diameter of 150 nm and a height of 60 nm while the less densely packed nanodisks (Fig 1 b and d) have a diameter of 140 nm and a height of 30 nm. The average (center-to-center) inter-particle distance is 250 nm for nanorings and 350 nm for nanodisks. For both sample structues, a 10 nm Ti adhesion layer was sputtered on top of a glass substrate, followed by a 100 nm Au layer and a 50 nm $SiO_2$ spacer layer. For the gold nanorings a self-assembled monolayer of 100 nm Polystyrene (PS) beads was deposited on top and subsequently covered with 30 nm of Au by sputter deposition. The nanorings are created by directional ion beam milling, opening up the top aperture of the nanorings arnd removing any gold in between the nanoparticles \cite{ref:18}. For the gold nanodisks a sacrificial PMMA layer is spincoated on top of the Au and $SiO_2$ layers. A self-assembled monolayer of 140 nm PS beads was deposited on top and serves as a template for the deposition of a 5 nm Au hard mask by evaporation. The beads are peeled off using tape and subsequently holes are formed in the PMMA layer below by RIE oxygen plasma etching. In the final steps, Au disks are deposited in the holes by evaporation and the PMMA resist is lifted off in acetone \cite{ref:19}. 

\begin{figure}[h]
\includegraphics[width=8.7cm]{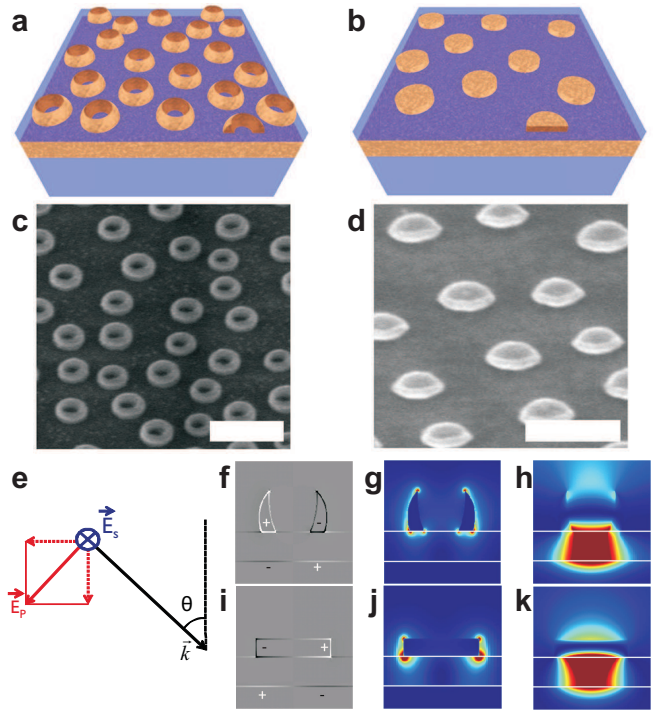}
\caption{Overview of the sample structure, measurement configuration and the involved plasmonic modes. (a) and (b) Schematic sample overview of a $1 \mu m^2$ area of rings and disks. (c) and (d) Scanning electron microscope pictures of the ring and disk samples. Scale bar 500nm. (e) Measurement configuration for the two polarization states. (f) and (i) Simulated charge plots at resonance for rings and disks. (g) and (j) Simulated electric field intensity at the electric dipole resonance for rings and disks. (h) and (k) Simulated induced magnetic dipole for rings and disks.}
\label{figure1}
\end{figure}

By scanning the angle of incidence, the electric dipole resonances (Fig 1 g and j) in the nanoparticles become spectrally detuned for both polarization states. Both for P- and S-polarized waves, the electric dipole in the nanoparticle couples to an electric quadrupole in the combined nanoparticle/gold film complex (Fig 1 f and i), which gives rise to an induced magnetic dipole (Fig 1 h and k) perpendicular to the electric dipole in the nanoparticle. 

The angle dependent spectroscopic ellipsometry measurements were performed using a commercial GESP5 \cite{ref:20} ellipsometer and a home-built setup based on a photo-elastic modulator (PEM) \cite{ref:21}. For the GESP5 setup, the polarization of the incident beam is modulated between P and S by a rotating polarizer, while for the PEM-based setup, the polarization is modulated between linear and left- and right- circular polarization states. Both measurement setups have different signal-to-noise ratios in different spectral ranges, so depending on the spectral position of the LSPR we choose the setup that performs best. All spectroscopic ellipsometry measurements shown here were performed with the GESP5 setup, except for the refractive index sensing measurements on gold rings in fig 5 b.

The phase information is extracted by performing lock-in measurements at the modulation frequency. The measured quantities $\tan(\Psi)$ and $\cos(\Delta)$ are related by the main equation of ellipsometry: 
\begin{equation}
\begin{split}
\rho = & \frac{R_{P}}{R_{S}} = \tan(\Psi) \exp(i \Delta) \\
  & = \tan(\Psi) (\cos(\Delta) + i \sin(\Delta))
\end{split}
\end{equation} 

and represent the amplitude reflection ratio between P and S $(\tan(\Psi))$ and the phase difference between the reflected signals $\Delta$ for the 2 polarizations (reflected in the $\cos(\Delta)$ value).  

In that way, the angle dependence of the plasmon resonances and the extremely narrow phase transitions at their central frequency could be probed experimentally. By properly designing the nanoparticle shape and density, the interparticle coupling can be tuned resulting in two spectrally slightly detuned resonances for P- and S-polarized incident light. An overview of the angle dependent measurements on the nanorings is given in figure 2. Panels a and b show the intensity based reflection spectra for P- and S-polarized incident waves, while panels c and d show the reflection ratio $\tan(\Psi)$ and the phase difference $\cos(\Delta)$ between the two polarization states. With increasing incident angle, the P-resonance shows a blue shift, while the S-resonance shows a red shift. These resonance shifts are reflected in the phase sensitive ellipsometry measurements, in which at the center frequency of the plasmon resonances a minimum and maximum in the reflection ratio (Fig 2 c), and a pronounced phase difference between P and S (Fig 2 d) is observed. 

\begin{figure}[h]
\includegraphics[width=8.7cm]{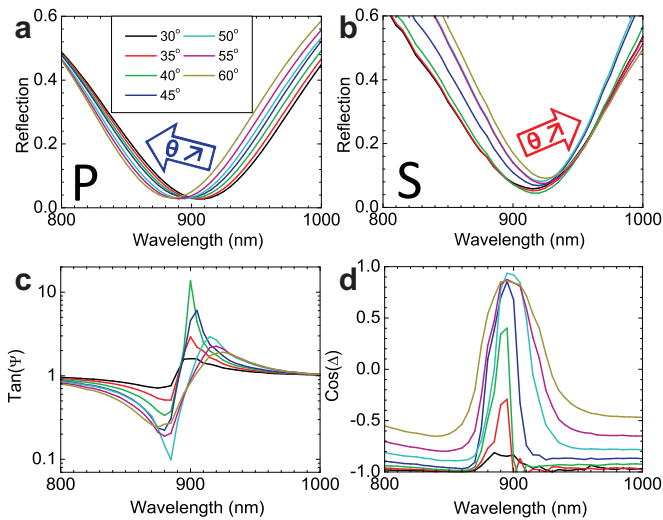}
\caption{Angle dependent measurement data for gold nanorings. (a) and (b) Measured intensity based reflection spectra for P- and S-polarizations. (c) Measured values of $\tan(\Psi)$, the amplitude reflection ratio between P- and S-polarizations.  (d) Measured values of $\cos(\Delta)$, with $\Delta$ the phase difference between both polarization states.}
\label{figure2}
\end{figure}

The angle- and polarization dependent optical response of our samples was simulated using Comsol multiphysics \cite{ref:22} (figure 3). The different spectra were calculated for a square lattice of rings with a pitch of 250 nm (matched to the experimental average inter-particle distance). By using periodic (Bloch) boundary conditions and scanning all the angles involved for both polarizations, the angle dependent reflection and the ellipsometric parameters were extracted. The reflected waves for P- and S-polarized waves were recorded and their electric and magnetic fields were averaged out over one unit cell, allowing to extract the amplitude and phase in order to evaluate the values of $\tan(\Psi)$ and $\cos(\Delta)$. A nice qualitative agreement is obtained, where the P-resonance and S-resonances show the blue shift (Fig 3 a) and red shift (Fig 3 b) respectively, similar to the experimental data. Contrary to the measured data, the minimum of the reflection dip shows a decrease with increasing angle of incidence. This behavior can be attributed to the random particle distribution in our samples, that gives rise to inhomogeneous broadening, which is most pronounced for large angles of incidence due to increasing spot sizes. This explains why the maxima and minima in the reflection ratio $(\tan(\Psi))$ are not observed at the largest angle of incidence in the experimental spectra, contrary to the simulated data (Fig 3 c). Interestingly, if we compare the magnitude of the phase difference between P and S, we clearly observe that the largest transitions are observed at the maxima and minima in the reflection ratio, which occur around $45^{o}$ in the experiments (Fig 2 d), and at $75^{o}$ in simulations (Fig 3 d). 

\begin{figure}[h]
\includegraphics[width=8.7cm]{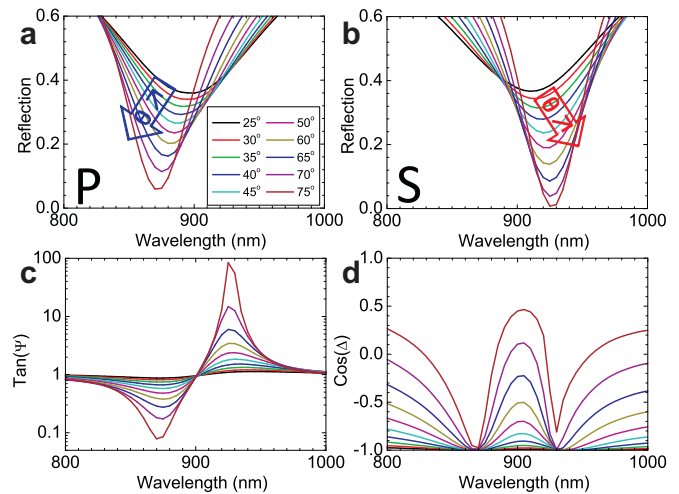}
\caption{Agle dependent simulation data for a periodic array of gold nanorings. (a) and (b) Simulated intensity based reflection spectra for P- and S-polarizations. (c) Simulated values of $\tan(\Psi)$, the reflection ratio between P- and S-polarizations.  (d) Simulated values of $\cos(\Delta)$, with $\Delta$ the phase difference between both polarization states.}
\label{figure2}
\end{figure}

Similar measurements were performed for gold nanodisks, which have comparable sizes to the rings, but are less densely packed due to different fabrication parameters. An overview of the angle dependent measurements for both polarizations is given in figure 4. Both for P- (Fig 4 a) and S-polarized (Fig 4 b) waves, the electric dipole resonances are blue shifted compared to the nanorings. With increasing angle of incidence, for the P-polarization, a minor blue shift is observed, while the S-resonance shows a pronounced red shift. Both resonances show much more spectral overlap compared to the nanorings, which results in a totally different behavior in their reflection ratio $\tan(\Psi)$ (Fig 4 c). For small angles of incidence we observe a maximum in the reflection ratio at shorter wavelengths, while at larger angles of incidence the S-resonance shifts to longer wavelengths with respect to the P-resonance, resulting in similar spectra as for gold nanorings. If we take a closer look at the phase difference between P and S (Fig 4 d), we observe again two phase jumps, one for each polarization state, which are smaller in magnitude compared to the nanorings. These smaller phase jumps can be attributed to the increased spectral overlap for the two polarization states, which results in a smaller overall phase difference. 

\begin{figure}[h]
\includegraphics[width=8.7cm]{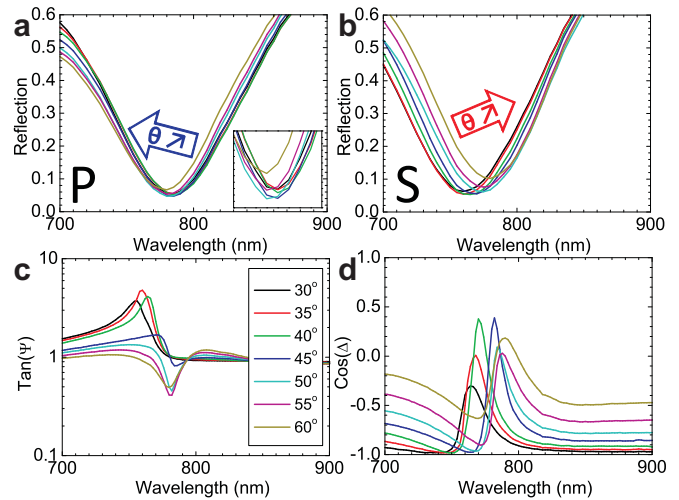}
\caption{Angle dependent measurements on gold nanodisks. Reflection spectra for P- (a) and S-polarization (b). (c) Reflection ratio $\tan(\Psi)$. (d) Phase difference $\cos(\Delta)$.}
\label{figure3}
\end{figure}

For the two nanoparticle geometries, we have illustrated that we can clearly identify the phase transitions at the LSPR frequency for the different polarization states. Now we want to take a closer look at the plasmon modes involved and the dominating inter-particle coupling mechanism. For both polarization states an electric dipole is excited in the nanoparticle (Figure 1 g and j), which shows a very broad linewidth (around 140 nm for rings and 100 nm for disks). This electric dipole resonance couples directly (P-polarization) or capacitively (S-polarization) to an electric quadrupole mode in the combined nanoparticle/gold film complex (Figure 1 f and i). The electric quadrupole mode also results in an induced magnetic dipole moment (Figure 1 h and k), which is aligned perpendicular to the electric dipole in the nanoparticles. By varying the angle of incidence and the polarization, both the coupling efficiency to the different modes and the phase retardation are scanned, resulting in spectral shifts of the plasmon resonances. The inter-particle coupling is mediated by electric dipole/quadrupole coupling and magnetic dipole coupling, both in the longitudinal and transversal direction. If the scattered fields of the nanoparticle are in phase (out of phase) with the incident wave in the neigboring particles, the local resonance will be enhanced (opposed) and show a blue (red) shift. 

\begin{figure}[h]
\includegraphics[width=8.7cm]{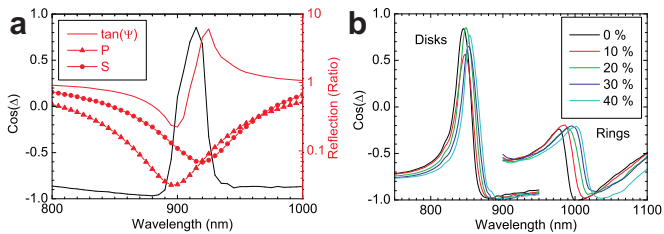}
\caption{Refractive index sensing measurements on gold nanodisks and nanorings for $45^{o}$ incidence. (a) Comparison of phase-sensitive measurements (black) and intensity based measurements (red) on gold nanorings in air, illustrating a dramatic line width reduction in the phase-difference between P and S. (b) Refractive index sensing measurements on disks and disks with variable concentrations of glycerol in water.}
\label{figure4}
\end{figure}

In a next step, the samples were mounted in a flowcell to perform refractive index sensing measurements with different concentrations of glycerol in water. An overview of the sensing measurements is presented in figure 5. In panel a, the linewidth reduction for the phase-sensitive measurements with respect to the intensity-based reflection measurements is clearly illustrated for nanorings in air. At the dip of the LSPR for both polarization states, a narrow phase transition is observed in $cos(\Delta)$. Panel b shows the wavelength shift as function of increasing glycerol concentrations in water. Here we used an incident angle of $70^{o}$, such that the incident angle at the sample/solution interface matches the $45^{o}$ incident angle of the reference measurement in air. As expected, with increasing refractive index, we observe a red-shift of both resonance positions. Different plasmon modes are excited for P- and S-polarization, which show different sensitivities to the refractive index. Moreover, depending on the spectral position of both resonances and their spectral overlap, the shape of the $cos(\Delta)$ signals can change dramatically for the different nanoparticles, which can be clearly observed when comparing the spectra for disks and rings. In order to calculate the figure of merit of the 2 model systems, we calculated the sensitivities for the combined phase-sensitive $cos(\Delta)$ signals of the P- and S-modes and compared these with the amplitude based reflection data. In the overview presented in table 1, we can clearly see that the phase sensitive measurements show much narrower line widths compared to their intensity-based counterparts. Due to this dramatic decrease in the spectral footprint, the FOM could be boosted 3.9 and 6.1 times up to 8.3 and 16.5 for nanodisks and nanorings respectively.  

\begin{center}
	\begin{table}
	  \caption{Comparison between intensity- and phase-sensitive reflection measurements on gold nanodisks and nanorings at an incidence angle of $45^{o}$.}
    \begin{tabular}{ | c | c | c |}
     \hline
     & Disks & Rings \\ \hline
     $d\lambda / dn$ (nm/RIU) & 208 & 380 \\ \hline
     FWHM reflection (nm) & 98 & 139 \\ 
     FOM reflection & 2.1 & 2.7 \\ \hline
     FWHM $cos(\Delta)$ (nm) & 25 & 23 \\
     FOM $cos(\Delta)$ & 8.3 & 16.5 \\ \hline
     FOM increase ratio & 3.9 & 6.1 \\ \hline
    \end{tabular}
    \end{table}
\end{center}

To summarize, we have shown that the FOM for reflection based refractive index sensing can be largely increased by measuring the phase of the reflected beam instead of its intensity only. Around the center frequency of the LSPR, the electron cloud makes the transition of in- to out-of-phase oscillation with respect to the driving field, which is an intrinsic property of a plasmon resonance. This phase jump shows a much smaller spectral footprint than the intensity based reflection measurements, resulting in much narrower line widths and largely increased values of the FOM. For the nanoparticles investigated in this work, we managed to increase the FOM up to 6 times for intrinsically broad dipole resonances. 

{\bf Acknowledgements: } The authors thank E. Vandenplas and J. Feyaerts for technical support. K.L. Acknowledges IWT Flanders and P.V.D. acknowledges FWO Flanders for financial support.

\end{document}